\documentclass[journal]{IEEEtran}
\IEEEoverridecommandlockouts
\usepackage{amsmath,amsfonts}
\usepackage{algorithmic}
\usepackage{array}
\usepackage[caption=false,font=footnotesize,labelfont=rm,textfont=rm]{subfig}
\usepackage{textcomp}
\usepackage{stfloats}
\usepackage{url}
\usepackage{verbatim}
\usepackage{graphicx}
\usepackage{balance}
\usepackage{xcolor}
\usepackage{epstopdf}
\usepackage{multirow}
\usepackage{booktabs}
\usepackage{latexsym}
\usepackage{cite}
\usepackage{algorithm}
\usepackage{hyperref}
\usepackage{marvosym}
\usepackage[numbers, sort&compress]{natbib}
\usepackage{nomencl}
\usepackage{textcomp,mathcomp}
\usepackage{color}
\usepackage{enumitem}
\usepackage{mathrsfs}
\usepackage{orcidlink}

\newenvironment{termtable}[1][2cm]{%
	\def\term##1##2{\item[$##1$] ##2}%
	\itemize[left=0pt .. #1, itemindent=0pt,
	align=parleft, nosep]
}
{%
	\enditemize
}
\makenomenclature

\begin{document}
	
	\title{Dynamic Virtual Power Plants with Robust Frequency Regulation Capability}

	\author{Xiang Zhu~\orcidlink{0009-0003-2742-3433},~\IEEEmembership{Graduate Student Member,~IEEE,}
	Hua Geng~\orcidlink{0000-0002-8336-6731},~\IEEEmembership{Fellow,~IEEE,}
    Hongyang Qing~\orcidlink{0000-0003-1280-3081},~\IEEEmembership{Member,~IEEE,}\\
    Grant Ruan~\orcidlink{0000-0003-2660-9298},~\IEEEmembership{Member,~IEEE,}
    and Xiuqiang He~\orcidlink{0000-0002-3755-7553},~\IEEEmembership{Member,~IEEE}

	\thanks{\textit{(Corresponding author: Hua Geng)}}	
	\thanks{X. Zhu, H. Geng, H. Qing, and X. He are with the Department of Automation, Tsinghua University, Beijing 10084, China and Beijing National Research Center for Information Science and Technology, Tsinghua University, Beijing 10084, China (email: zhu-x22@mails.tsinghua.edu.cn; genghua@tsinghua.edu.cn; qinghy@mail.tsinghua.edu.cn; hxq19@tsinghua.org.cn).}
	\thanks{G. Ruan is with the Laboratory for Information \& Decision Systems (LIDS), MIT, Boston, MA 02139 (email: gruan@mit.edu).}
	\thanks{This work was supported by Science and Technology Project of State Grid Corporation of China (No. 5100-202456019A-1-1-ZN).}		
}
	
	\maketitle
	
	\begin{abstract}  	

    The rapid integration of inverter-based resources (IBRs) into power systems has identified frequency security challenges due to reduced inertia and increased load volatility. This paper proposes a robust power reserve decision-making approach for dynamic virtual power plants (DVPPs) to address these challenges, especially under temporally sequential and uncertain disturbances. An analytical model is developed to characterize the system's frequency response dynamics, enabling the quantification of virtual inertia and virtual damping requirements to meet rate-of-change-of-frequency (RoCoF), frequency nadir, and steady-state deviation constraints. By analytically deriving the regulation power dynamics, the required virtual inertia and damping parameters for the DVPP are determined in a robust way. Then, the total power reserve decision is made by optimally allocating the parameters and calculating the actual power reserves for IBRs, fully considering their economic diversity. Finally, case studies conducted on an IEEE nine-bus system demonstrate the effectiveness of the proposed approach. The results indicate the high reliability of the proposed approach in ensuring frequency security.
    
	\end{abstract}
	
	
	\IEEEpeerreviewmaketitle
	\section*{Nomenclature}
	
	\subsection*{\bf{Abbreviations}}
	\begin{termtable}[1.9cm]
		\term{\text{SG}}{Synchronous generator}
        \term{\text{PV}}{Photovoltaic}
		\term{\text{IBR}}{Inverter-based resource}
        \term{\text{FFR}}{Fast frequency response}
        \term{\text{PFR}}{Primary frequency regulation}
		\term{\text{DVPP}}{Dynamic virtual power plant}
        \term{\text{RoCoF}}{Rate of the change of frequency}
	\end{termtable}
	
	\subsection*{\bf{Parameters}}
	\begin{termtable}[1.9cm]
		\term{\Delta P}{The amplitude of power disturbance [MW]}
		\term{D_{0}}{The load damping coefficient [MW/Hz]}
		\term{H_{0}}{The inertia parameter of SGs [MWs/Hz]}
		\term{R}{The droop coefficient of SGs [MW/Hz]}
		\term{T^\text{SG}}{The PFR response time of SGs [s]}	
		\term{\Delta f_\text{lim}^\text{RoCoF}}{RoCoF limitation of frequency [Hz/s]}
		\term{\Delta f_\text{lim}^\text{Nadir}}{Nadir limitation of frequency [Hz]}
        \term{\Delta f_\text{lim}^\text{ss}}{Steady-state limitation of frequency [Hz]}
		\term{P^\text{av}_{i}}{Upper limit of $i$th IBR's power injection [MW]}
		\term{N}{The number of IBRs involved in the DVPP}
        \term{n}{The number of periods of sequential disturbances}
        \term{\tau}{The disturbance prediction window length [s]}
        \term{\mathcal{P}_{i}}{The probabilistic factors of disturbances}
	\end{termtable}
	
	\subsection*{\bf{Variables}}
	\begin{termtable}[1.9cm]
		\term{\Delta f(t)}{The system frequency response [Hz]}
        \term{\Delta P^\text{Inj}}{The active power injection of DVPP [MW]}
		\term{R^{U}}{The upward power reserve for DVPP [MW]}
		\term{R^{D}}{The downward power reserve for DVPP [MW]}
		\term{\Delta P^\text{IBR}_{i}(t)}{The active power injection of $i$th IBR [MW]}
		\term{H_\text{DVPP}}{Virtual inertia of the DVPP [MWs/Hz]}
		\term{D_\text{DVPP}}{Virtual damping of the DVPP [MW/Hz]}
		\term{H_{i}}{Virtual inertia of the $i$th IBR [MWs/Hz]}
		\term{D_{i}}{Virtual damping of the $i$th IBR [MW/Hz]}
        
	\end{termtable}
	
	\section{Introduction}

    \IEEEPARstart{F}{uture} power systems are undergoing a significant transformation, with inverter-based resources (IBRs) progressively becoming dominant in the energy landscape~\cite{IBR-TIA1, zhu-TIA}. Traditional synchronous generators (SGs), renowned for their high rotational inertia, are being phased out by IBRs. This transition leads to a substantial decline in system inertia, consequently amplifying both the rate of change of frequency (RoCoF) and the magnitude of frequency deviations in power grids~\cite{IBR-TIA2, Back-1, Review-He, IBR-TIA3}. Furthermore, the widespread adoption of emerging loads (e.g., electric vehicles and smart buildings) and the growing demand for flexibility through demand response programs have exacerbated load curve volatility~\cite{ADD-1}. These factors complicate active power balancing and pose critical challenges to frequency security, particularly by causing time-sequential and uncertain frequency deviations~\cite{LoadFlu-1, LoadFlu-2}. 

    A widely adopted strategy to address these challenges is to utilize IBRs for delivering ancillary services, such as virtual inertia support and primary frequency response, thereby offsetting the diminishing role of SGs~\cite{DER}. To enhance IBR coordination in frequency support, the concept of dynamic virtual power plants (DVPPs) has emerged~\cite{VPP-ruan, Review-DVPP, zhu2025optimal}. A DVPP aggregates heterogeneous IBRs via contractual coordination, enabling them to collectively mimic the inertial and damping characteristics of conventional SGs. Beyond traditional static setpoint scheduling, the core principle of a DVPP lies in coordinating IBRs through business contracts to fulfill dynamic grid requirements~\cite{DVPP1}. This paradigm enables small-scale IBRs to collaboratively engage in dynamic ancillary services, including real-time frequency regulation~\cite{zhu2024dynamic}.

    Recent research on DVPP control strategies has primarily focused on localized coordination mechanisms. For instance, \cite{DVPP1} proposed an adaptive divide-and-conquer strategy to deliver tailored dynamic ancillary services. Building on this, \cite{DVPP2} investigated the spatio-temporal characteristics of ancillary services and developed grid-forming control with distributed architectures for DVPPs. To optimize signal distribution within DVPPs, \cite{DVPP3} introduced a measurement-based method for dynamically allocating regulation signals among IBRs. Moreover, \cite{DVPP4} designed a decentralized closed-loop control scheme to maximize the cooperative capabilities of DVPPs in ancillary service provision.

    On top of DVPP control strategies, the power reserve discussion of DVPP is also a critical issue. Virtual inertia/damping scheduling schemes present a promising solution to the power reserve decision by quantifying power dynamics and optimizing the fast frequency response~(FFR) participation for IBRs~\cite{COM1, COM2, COM3}. For example, \cite{COM1, COM2} established a dynamic security-constrained framework to quantify inertia requirements in IBR-dominated systems, enabling cost-effective operational strategies. Extending this, \cite{COM3} incorporated both inertia and damping requirements to guide IBR reserve allocation, while~\cite{COM4} generalized the framework to hybrid systems with fast and slow generators.

    Despite these advancements, a critical research gap persists: existing methods inadequately assess frequency regulation requirements under temporally sequential and uncertain active power disturbances. Such disturbances induce frequency deviations via the active power-frequency coupling mechanism~\cite{VPP1}. Current approaches~\cite{COM1, COM2, COM3, COM4} often oversimplify temporally sequential disturbances by averaging them as single-step events with fixed magnitudes, assuming temporal independence and starting from the nominal point. However, practical scenarios involve interdependent disturbances—for instance, a load increase followed by a decrease may partially cancel each other, whereas consecutive disturbances in the same direction amplify frequency deviations. This temporal dependency invalidates traditional step-disturbance assumptions and complicates frequency security analysis. Furthermore, simplified models fail to ensure safe dynamic performance, especially as temporally sequential and uncertain disturbances become prevalent in grids with volatile renewable generation and flexible loads~\cite{Flu1, Flu2}. The cumulative effects of sequential disturbances further obscure frequency response analysis and power reserve decisions. 

\begin{table*}[htbp]
  \centering
  \caption{Comparison with earlier works}
    \begin{tabular}{ccccccc}
    \toprule
    \multirow{2}[4]{*}{Ref.} & \multirow{2}[4]{*}{Multi disturbance consideration} & \multicolumn{3}{p{17.275em}}{Time-domain metrics of frequency security} & \multirow{2}[4]{*}{Power reserve calculation} & \multirow{2}[4]{*}{Power reserve allocation optimum} \\
\cmidrule{3-5}    \multicolumn{1}{c}{} & \multicolumn{1}{c}{} & RoCoF & nadir & steady-state deviation & \multicolumn{1}{c}{} & \multicolumn{1}{c}{} \\
    \midrule
    \cite{COM1}     &      & \checkmark     & \checkmark     & \checkmark     & \checkmark     & \checkmark \\
    \cite{COM2}     &      & \checkmark     & \checkmark     & \checkmark     & \checkmark     & \checkmark \\
    \cite{COM3}     &      & \checkmark     & \checkmark     & \checkmark     & \checkmark     & \checkmark \\
    \cite{FFR3}     &      & \checkmark     & \checkmark     & \checkmark     &      &  \\
    \cite{ADD-4}    &      &      &      &      & \checkmark     &  \\
    \cite{ADD-5}    &      & \checkmark     & \checkmark     & \checkmark     & \checkmark     & \checkmark \\
    \cite{ADD-6}    &      & \checkmark     & \checkmark     & \checkmark     &      &  \\
    \cite{Flu1}     & \checkmark     & \checkmark     &      &      & \checkmark     & \checkmark \\
    \cite{Flu2}     & \checkmark     & \checkmark     &      &      & \checkmark     & \checkmark \\
    Proposed        & \checkmark     & \checkmark     & \checkmark     & \checkmark     & \checkmark     & \checkmark \\
    \bottomrule
    \end{tabular}%
  \label{tab:earlier}%
\end{table*}%

    To bridge these gaps, this paper proposes a novel approach for power reserve decision-making under temporally sequential and uncertain disturbances, as shown in Fig.~\ref{fig.framework}. First, we develop analytical models to characterize the closed-form dynamic relationship between DVPP-level frequency regulation parameters (i.e., virtual inertia and damping) and system performance. On this basis, the optimal power reserve requirements are derived for the DVPP using an economic, diversity-aware optimization scheme. The comparison with earlier works is presented in TABLE~\ref{tab:earlier}.
	
	The major contributions are twofold:
	\begin{enumerate}

        \item  \textbf{Closed-form dynamics characterization: }This work develops separate analytical models to capture the temporally coupled dynamics of system frequency response and regulation power in DVPPs under sequential disturbances with uncertain occurrence times. Through closed-form derivations, we explicitly characterize the influence of tunable IBR parameters (virtual inertia and damping), aggregated by the DVPP, on both frequency security and active power injection patterns.

        \item \textbf{Robust parameter optimization scheme: } A robust feasible region for the tunable parameters is formulated, ensuring compliance with RoCoF, nadir, and steady-state constraints under worst-case disturbances. This region aids in determining the required virtual inertia and damping for the DVPP. The total power reserve decision is made by optimally allocating parameters and calculating the actual power reserves for IBRs, fully accounting for their economic diversity.
        
	\end{enumerate}
	
    The remainder of this paper is structured as follows: Section~\ref{sec2} formalizes the problem of frequency regulation under temporally sequential and uncertain disturbances. Section~\ref{sec3} develops the closed-form dynamic model linking DVPP control parameters to frequency response and power disturbance. Section~\ref{sec4} shows how to select optimal parameters for power reserve decision-making. Section~\ref{sec5} validates the approach through case studies on a modified IEEE nine-bus system. Finally, Section~\ref{sec6} concludes this paper.
	
	\section{Problem Formulation} \label{sec2}

	\begin{figure}
		\centering
		\includegraphics[width=0.95\linewidth]{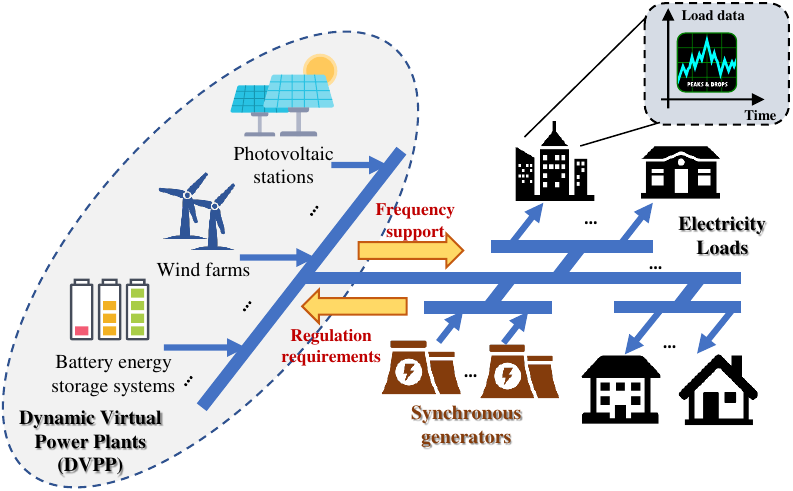}
		\caption{A DVPP interacting with the grid. It provides sufficient frequency response to the regulation requirements from the grid side.} 
		\label{fig.framework} 
	\end{figure}
	
	\begin{figure}
		\centering
		\includegraphics[width=0.95\linewidth]{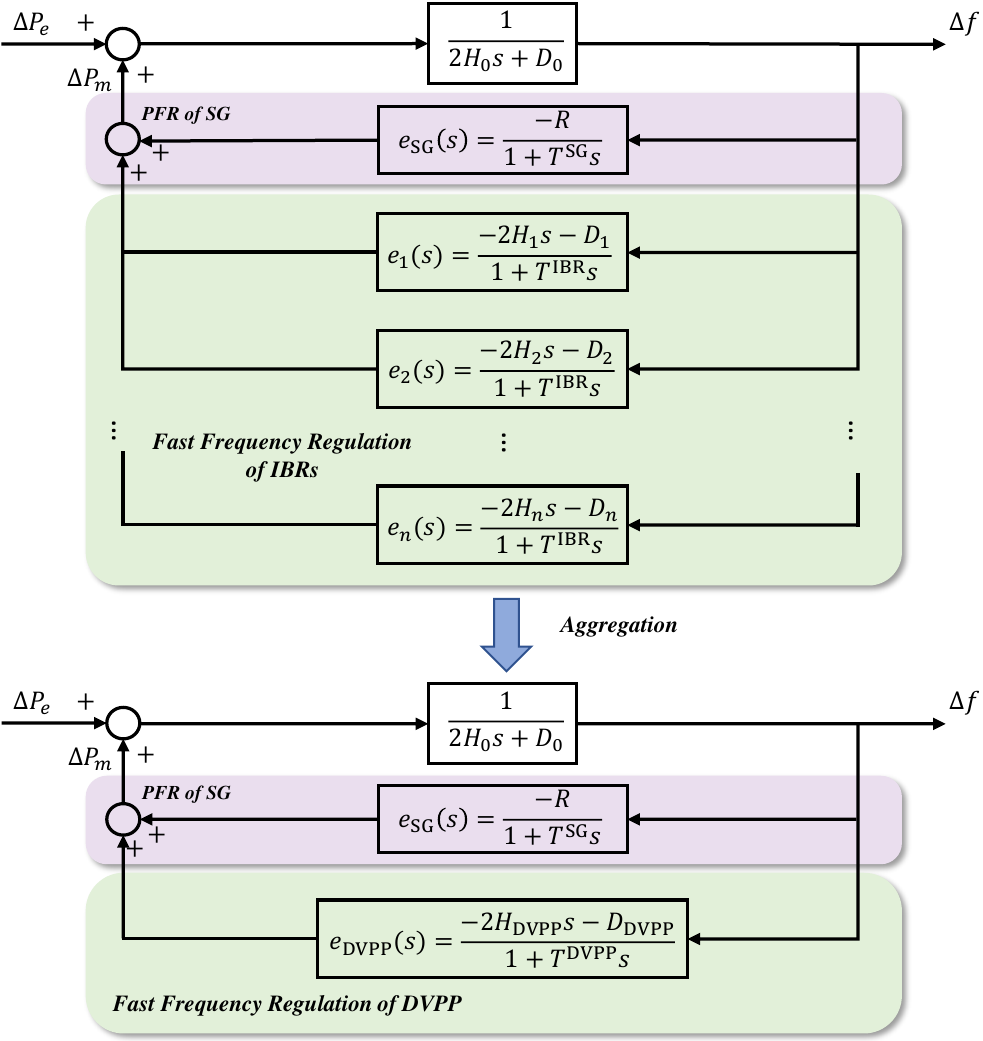}
		\caption{The block diagram illustrates the system frequency response. The forward path represents the inertia and damping of the grid. The purple and green parts describe the PFR of SGs and the frequency responses of DVPP, respectively.} 
		\label{fig.block} 
	\end{figure}
	
	The dynamic virtual power plant~(DVPP) is a set of IBRs that can be coordinated and controlled for dynamic ancillary services, such as fast frequency response. In this paper, we consider a DVPP containing various flexible IBRs, e.g., photovoltaic~(PV) generation systems, wind turbines and battery energy storage systems, to provide both commercial energy trading and frequency regulation as shown in Fig. \ref{fig.framework}. The DVPP, by aggregating IBRs, has available power capacity. It can determine a certain amount of power for local grid frequency regulation, while the remaining power is sold on the electricity market to generate economic profit. This means that the power reserve allocated by the DVPP for frequency regulation should be minimized as much as possible to enhance economic efficiency while still meeting the frequency security requirements. Additionally, it should be clarified that in this study, the uncertainty of wind and PV power generation is not considered, as this uncertainty can be mitigated by the presence of the battery energy storage system~\cite{Uncertainty-1}.

	\begin{figure}
		\centering
		\includegraphics[width=0.95\linewidth]{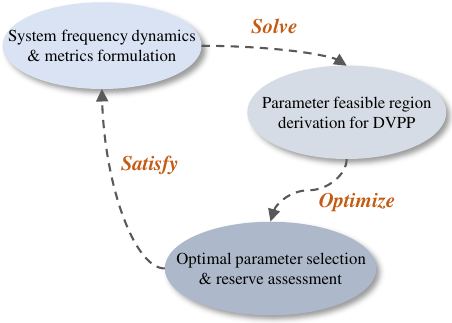}
		\caption{The diagram of the proposed power reserve decision-making framework.} 
		\label{fig.logic} 
	\end{figure} 

    Temporally sequential and uncertain active power disturbances are inevitable in power systems with new types of loads, leading to continuous frequency fluctuations over short periods. To address the frequency regulation issue under such disturbance scenarios, the DVPP is configured to predict the potential disturbance curves within a fixed-length time window in the future. This enables the optimal determination of the frequency regulation strategy (i.e., active power reserves) for the DVPP. From a physical perspective, the DVPP aggregates numerous IBRs operating with virtual inertia and droop control schemes to provide the required active power injections when power imbalances occur~\cite{FFR2, FFR4}. The performance of the DVPP's FFR control is determined by the equivalent parameters, $H_\text{DVPP}$ and $D_\text{DVPP}$, which are obtained by aggregating the specific control parameters of the power control loop of each IBR, i.e., $H_\text{DVPP} = \sum_{i=1}^{n} H_{i}$ and $D_\text{DVPP} = \sum_{i=1}^{n} D_{i}$. The aggregation relationship between the DVPP and IBRs is visually depicted in Fig. \ref{fig.block}. It should be noted that the focus of this paper is specifically on the transient dynamics of inertia response and primary frequency regulation (PFR) within the timescale of minutes. As such, secondary frequency regulation (SFR) and other controls operating over longer timescales are intentionally excluded for the sake of simplification.

    In summary, determining the power reserves required for frequency regulation hinges on identifying the aggregated virtual inertia and damping parameters at the DVPP level and subsequently allocating them optimally to the IBRs. This paper first models and conducts a dynamic analysis of the system's continuously fluctuating frequency. It then defines the feasible parameter region based on predefined frequency security metrics and identifies the optimal parameter allocation within this region to assess the required power reserves. The logic diagram of the proposed power reserve decision-making framework is shown in Fig.~\ref{fig.logic}.
    
	\section{The Robust Frequency Regulation Requirements Derivation} \label{sec3}
	
	\subsection{System Frequency Response under Single-Step Active Power Disturbance}
	
	For the sake of analysis, we first consider the frequency deviation under a single-step disturbance to derive the system frequency response dynamics. The frequency-domain representations are shown in~(\ref{fre_1}) based on the block diagram after aggregation in Fig. \ref{fig.block}. 
	\begin{equation}
		\left\{ \begin{aligned}
			& F(s)=\left( \Delta {{P}_{e}}(s)+\Delta {{P}_{m}}(s) \right)\frac{1}{2{{H}_{0}}s+{{D}_{0}}} \\ 
			& \Delta {{P}_{m}}=-\left( \frac{R}{1+{{T}^\text{SG}}s}+2{{H}_\text{DVPP}}s+{{D}_\text{DVPP}} \right)\Delta f(s)\\ 
		\end{aligned} \right.
		\label{fre_1}
	\end{equation}
	where $\Delta P_{e}(s)$ is considered as a step disturbance $\Delta P/s$, $R$ is the droop coefficient of SGs, $H_{0}$ and $H_\text{DVPP}$ are the inertia and virtual inertia parameters of SGs and DVPP, $D_\text{DVPP}$ is the virtual damping parameter of the DVPP, $D_{0}$ is the load damping coefficient. 
    
    By reconstruction, the frequency dynamics can be reformulated as in~(\ref{fre_re}), where $H = H_{0}+H_\text{DVPP}$ and $D = D_{0}+D_\text{DVPP}$. According to the characteristics of the second-order system, the small-signal stability of the closed-loop system is ensured under the scheduling of parameters $H_\text{DVPP}$ and $D_\text{DVPP}$.

    \begin{equation}
        \Delta f(s) = {P_e}(s) \cdot \frac{{1 + {T^{{\rm{SG}}}}s}}{{\left( {2Hs + D} \right)\left( {1 + {T^{{\rm{SG}}}}s} \right) + R}}
        \label{fre_re}
    \end{equation}
    
    Based on (\ref{fre_re}), the time-domain representation of system frequency under a single disturbance can be derived through the Inverse Fourier transform~($ \mathscr{L}^{-1}(\cdot)$) on $F(s)$ in (\ref{fre_2}).
	\begin{equation}
		\begin{aligned}
			& F(\Delta P,t) = \mathscr{L}^{-1} \left( \frac{{\Delta P}}{s} \cdot \frac{{(1 + T^\text{SG}s)}}{{(2Hs + D)(1 + T^\text{SG}s) + R}}
			\right) \\
			& = \frac{{\Delta P}}{{D + R}} \cdot [1 + {e^{ - \zeta {\omega _n}t}}\eta \sin ({\omega _d}t + \varphi )]
		\end{aligned}
		\label{fre_2}
	\end{equation}
	where 
	\begin{equation}
		H=H_{0}+H_\text{DVPP}, D=D_{0}+D_\text{DVPP}
	\end{equation}
	\begin{equation}
		\zeta  = \frac{{2H + D{T^{{\rm{SG}}}}}}{{2\sqrt {2{T^{{\rm{SG}}}}H(R + D)} }}
		\label{zeta}
	\end{equation}
	\begin{equation}
		{\omega _n} = \sqrt {\frac{{{D_{{\rm{VPP}}}} + {D_0} + R}}{{2H{T^{{\rm{SG}}}}}}} 
		\label{omega}
	\end{equation}
	\begin{equation}
		\eta  = \sqrt {\frac{{1 - 2{T^{{\rm{SG}}}}{\omega _n}\zeta  + {T^{{\rm{S}}{{\rm{G}}^2}}}\omega _n^2}}{{1 - {\zeta ^2}}}} 
		\label{eta}
	\end{equation}
	\begin{equation}
		\varphi  = \arctan (\frac{{{\omega _d}}}{{ - {T^{{\rm{SG}}}}\omega _n^2 + \zeta {\omega _n}}})
		\label{phi}
	\end{equation}
	
	\subsection{System Frequency Response Modeling under Temporally Sequential and Uncertain  Disturbances}

	\begin{figure}
		\centering
		\includegraphics[width=0.95\linewidth]{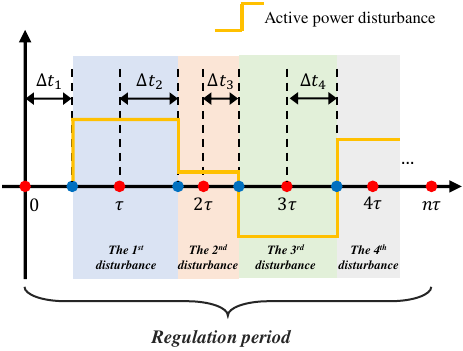}
		\caption{The diagram illustrates the fluctuations of the temporally sequential and uncertain active power disturbances during the presupposed regulation period $n\tau$.} 
		\label{fig.timeaxis}
	\end{figure}
    
    The equations in Eq. (\ref{fre_2})-(\ref{phi}) clearly describe the system's frequency response ($F(t)$) under a single-step active power disturbance ($\Delta P/s$). While the single-disturbance analysis offers advantages for deriving closed-form solutions, this assumption is overly restrictive and does not accurately reflect real-world operating conditions, where disturbances occur probabilistically. This section considers the uncertain occurrence of active power disturbances, as illustrated in Fig.~\ref{fig.timeaxis}. The total active power disturbance over a regulation period of $n\tau$ consists of $n$ segments, each of which may contain one disturbance. For the sake of simplifying the modeling process, the following assumptions are made in this study:
    \begin{itemize}
        \item The probability, magnitude, and direction of each disturbance are assumed to be known, based on accurate forecasts of system load and power generation.
        \item The potential occurrence times of disturbances are evenly distributed, with the interval between disturbances being greater than the system's response time to reach a quasi-steady state after each disturbance.
    \end{itemize}
    
    According to the frequency-disturbance causality, the frequency response under active power disturbances with probabilistic occurrence exhibits the same fluctuation tendency and continuity in the time domain. Thus, the system frequency is modeled using piecewise functions as described in (\ref{fre-multi1})-(\ref{fre-multi2}). Since the occurrence of the active power disturbances is probabilistic, the resulting frequency deviations also occur with the same probability. In the model, we introduce probabilistic factors~($0\le \mathcal P_{i} \le 1, i=1,2,...,n$) to capture the frequency response characteristics over continuous periods.
	\begin{equation}
		\Delta f(t) = \left\{ \begin{aligned}
			& \Delta {f_1}(t),\Delta t_{1} \le t \le \tau + \Delta t_{2} \\ 
			& \Delta {f_2}(t),\tau + \Delta t_{2} < t \le 2\tau+\Delta t_{3} \\
			& \vdots \\
			& \Delta {f_n}(t),(n-1)\tau+\Delta t_{n} < t \le n\tau\\
		\end{aligned} \right.
		\label{fre-multi1}
	\end{equation}
	where
	\begin{equation} 
		\left\{ \begin{aligned}
			& \Delta {f_1}(t) = \mathcal{P}_{1}\cdot F\left[ \Delta P_{1},t-\Delta t_{1} \right] \\
			& \Delta {f_2}(t) = \mathcal{P}_{2}\cdot F\left[\Delta P_{2},t - (\tau+\Delta t_{2}) \right] + \Delta {f_1}(\tau+\Delta t_{2})\\
			& \vdots\\
			& \Delta {f_n}(t) = \mathcal{P}_{n}\cdot F\left[\Delta P_{n},t-((n-1)\tau+\Delta t_{n}) \right] \\
            & + \Delta {f_{n - 1}}\left[ (n-1)\tau+\Delta t_{n} \right]
		\end{aligned} \right.
		\label{fre-multi2}
	\end{equation}
    where $\Delta f_{i}(t), i=1,2,...,n$ means the possible frequency response under active power disturbances. The disturbance prediction window length is set as $\tau$ and thus the regulation period is $n\tau$. While $\Delta t_{i}, i=1,2,...,n$ indicates the time point at the occurrence of the $i$th disturbance. Besides, $\mathcal{P}_{i}\cdot F\left[\Delta P_{i},t-((i-1)\tau+\Delta t_{i}) \right], i=1,2,...,n$ means that frequency response caused by the $i$th active power disturbance $\Delta P_{i}$ with a probability of $\mathcal P_{i}$. 
    
    Additionally, it should be noted that the active power disturbances considered in this study exhibit dual uncertainty: whether a disturbance will occur within each period of length $\tau$, and if it does, the specific moment at which it will occur. The uncertainty in disturbance magnitudes is not considered according to the earlier assumptions.
    
	\subsection{The Worst Disturbance Decision-Making}

    To address the uncertainty of active power disturbances, a worst disturbance decision-making scheme is introduced in this section. Initially, the aggregated virtual inertia ($H_\text{DVPP}$) and virtual damping ($D_\text{DVPP}$) are set to their respective initial values. Using these parameters, the possible system frequency responses under various disturbances are analyzed to identify the worst disturbances. The worst disturbances are evaluated using time-domain security metrics, specifically the rate of change of frequency (RoCoF), frequency nadir and steady-state deviations. Once the frequency regulation requirements for the worst disturbances are satisfied by the DVPP, the requirements for all other disturbances are also covered.

    The RoCoF metric is used to assess the rate of change of the system frequency, i.e., $\left| \Delta \dot f(t) \right|$, following the occurrence of disturbances. Since the RoCoF is maximized at the moment of disturbance occurrence, the metric~($M^{R}$) is simplified to the time point of disturbance occurrence~(\ref{metric-1})-(\ref{metric-2}).
    \begin{equation}
        M^{R} = \text{max}\left| \Delta \dot f(t^{R}_{i}) \right|,\ i = 1,2,...,n
        \label{metric-1}
    \end{equation}
    where
	\begin{equation}
		\left| {\Delta \dot f(t^{R}_{i})} \right| = \left| {\frac{{\Delta {P_i}}}{{2({H_{{\rm{DVPP}}}} + {H_0})}}} \right|,\ i = 1,2,...,n
	\end{equation}
    \begin{equation}
        t^{R}_{i} = (i-1)\tau + \Delta t_{i},\ i = 1,2,...,n 
        \label{metric-2}
    \end{equation}
    
    The frequency nadir metric~($M^{N}$) is used to evaluate the maximum degree of the system frequency deviation~(\ref{metric-3}). Considering the sequential relevance of frequency response, $M^{N}$ is determined by the time-domain features of the active power disturbance.
    \begin{equation}
        M^{N} = \text{max}\left| \Delta f(t) \right|
        \label{metric-3}
    \end{equation}

    Within the frequency regulation period, the potential time points at which the frequency nadir may occur include the start/end time points and the time of the frequency response peak ($t^{P}_{i}$). The time of the frequency peak can be derived analytically, with the result shown in (\ref{peaktime-1}), which satisfies the extreme point condition~(\ref{peaktime-2}). Thus, the nadir value metric can be expanded as in~(\ref{peaktime-3}).
    \begin{equation}
        t^{P}_{i} = t^{R}_{i}+\frac{\arctan (\frac{{{\omega _d}}}{{\zeta  \cdot {\omega _n}}}) - \varphi }{\omega_{d}}
        \label{peaktime-1}
    \end{equation}
    \begin{equation}
        {\left. {\frac{{{\rm{d}}\Delta f(t)}}{{{\rm{d}}t}}} \right|_{t = {t^{P}_{i}}}} = 0
        \label{peaktime-2}
    \end{equation}
    \begin{equation}
        M^{N} = \text{max} \left\{ \left| \Delta f(t^{P}_{i})\right|, \left| \Delta f(\Delta t_{1})\right|, \left| \Delta f(n\tau)\right| \right\}
        \label{peaktime-3}
    \end{equation}

    The steady-state metric ($M^{S}$) is derived in (\ref{Steady-state})-(\ref{Steady-state-2}) based on system frequency formulation~(\ref{fre-multi1})-(\ref{fre-multi2}).
    \begin{equation}
        M^{S} = \text{max}\left| \Delta f^{ss}_{i} \right|,\ i=1,2,...,n
        \label{Steady-state}
    \end{equation}
    \begin{equation}
        \left\{ \begin{aligned}
			& \Delta f^{ss}_{1}(t) = \Delta f_{1}(\tau+\Delta t_{2}) \\
			& \Delta f^{ss}_{2}(t) = \Delta f_{1}(\tau+\Delta t_{2})+\Delta f_{2}(2\tau+\Delta t_{3}) \\
			& \vdots\\
			& \Delta f^{ss}_{n}(t) = \Delta f_{1}(\tau+\Delta t_{2})+...+\Delta f_{n}(n\tau)
		\end{aligned} \right.
        \label{Steady-state-2}
    \end{equation}
    
    The worst disturbances are defined as one~($L^{w}$) that causes the maximum RoCoF metric~($M^{R}$), nadir metric~($M^{N}$) and steady-state metric~($M^{S}$). Based on the dynamic characteristics of the system frequency formulated in~(\ref{fre_2}), both the maximum RoCoF and nadir values are triggered by disturbances with the largest magnitude. And the maximum steady-state deviation results from the occurrence of all possible disturbances. Therefore, the worst disturbances refer to cases where all possible disturbances occur, including those with the maximum possible magnitude.
    
    \subsection{Robust Feasible Parameter Region Formulation}

    The DVPP can provide sufficient support by tuning proper parameters, i.e., aggregated virtual inertia~($H_\text{DVPP}$) and virtual damping~($D_\text{DVPP}$). In this section, we derive a robust feasible parameter region for ($H_\text{DVPP}, D_\text{DVPP}$), which satisfies the constraints on the RoCoF, the nadir value and the steady-state value of frequency under $L^{w}$. Specifically, the constraint on the RoCoF focuses on the time points when the power disturbances occur, i.e., $t^{R}_{i}$ in~(\ref{metric-2}), and is formulated as in~(\ref{metric-rocof}). 
	\begin{equation}
		\left. M^{R}\right|_{L^{w}}  \le \left| {\Delta f_{\lim }^\text{RoCoF}} \right|
		\label{metric-rocof}
	\end{equation}

	The constraint on the frequency nadir value~(\ref{metric-nadir}) refers to the frequency response's maximum or minimum values triggered sequentially and is formulated earlier in (\ref{peaktime-3}). 
	\begin{equation}
		\left. M^{N}\right|_{L^{w}} \le \left| {\Delta f_{\lim }^\text{Nadir}} \right|
		\label{metric-nadir}
	\end{equation}

	The constraint on the steady-state of system frequency is formulated in~(\ref{metric-qss}) and the steady-state value of~(\ref{fre-multi1}) is also described by a piecewise function in (\ref{Steady-state})-(\ref{Steady-state-2}).  
	\begin{equation}
		\left. M^{S}\right|_{L^{w}} \le \left| {\Delta f_{\lim }^\text{ss}} \right|
		\label{metric-qss}
	\end{equation}

    From what has been discussed above, the robust feasible parameter region~($\Omega$) is formulated as in (\ref{safety-region}) that represents the feasible range of the aggregated virtual inertia~($H_\text{DVPP}$) and virtual damping~($D_\text{DVPP}$) under the worst disturbances $L^{w}$.
	\begin{equation}
		\Omega(H_\text{DVPP},D_\text{DVPP}) = \left\{ \begin{aligned}
			& \left. M^{R}\right|_{L^{w}}  \le \left| {\Delta f_{\lim }^\text{RoCoF}} \right|\\
			& \left. M^{N}\right|_{L^{w}} \le \left| {\Delta f_{\lim }^\text{Nadir}} \right|\\
            & \left. M^{S}\right|_{L^{w}} \le \left| {\Delta f_{\lim }^\text{ss}} \right|\\
			% & H_\text{DVPP} \le H_\text{DVPP}^{max}\\
			% & D_\text{DVPP} \le D_\text{DVPP}^{max}\\
		\end{aligned} \right.
		\label{safety-region}
	\end{equation}
	
	\section{Optimal Power Reserve Assessment for DVPP} \label{sec4}

    After determining the robust feasible parameter region $\Omega$ for aggregated virtual inertia~($H_\text{DVPP}$) and aggregated virtual damping~($D_\text{DVPP}$), an optimal parameter selection becomes imperative since the feasible domain contains infinite parameter combinations. This section presents a two-stage optimization strategy to systematically determine the optimal DVPP parameters. On this basis, the economic differences of each IBR are comprehensively considered for parameter allocation optimization. Each IBR calculates its upward and downward reserve requirements based on the allocation results, and the final upward and downward reserve demands of the DVPP are obtained by summing the reserves of each IBR.
    
	\subsection{Active Power Injections Derivation}	

	For the analytic derivation of the power injections of the DVPP, we first derive its time-domain representation under single-step disturbance ${\Delta P}/s$ in (\ref{power-1})-(\ref{power-2}), where the negative sign indicates the frequency support effect.	
	\begin{equation}
		I(\Delta P,t) =  - \frac{{A + {e^{ - \zeta {\omega _n}t}}\left( {\frac{{C - \zeta {\omega _n}B}}{{{\omega _d}}}\sin ({\omega _d}t) + B\cos ({\omega _d}t)} \right)}}{{2H{T^{{\rm{SG}}}}}}
		\label{power-1}
	\end{equation}
	where
	\begin{equation}
		\left\{ \begin{aligned}
			& A = \frac{{\Delta P \cdot {D_{{\rm{DVPP}}}}}}{{\omega _n^2}}\\
			& B = 2{T^{{\rm{SG}}}}{H_{{\rm{DVPP}}}}\Delta P - \frac{{\Delta P \cdot {D_{{\rm{DVPP}}}}}}{{\omega _n^2}}\\
			& C = ({T^{{\rm{SG}}}}{D_{{\rm{DVPP}}}} + 2{H_{{\rm{DVPP}}}})\Delta P - \frac{{2\zeta \Delta P \cdot {D_{{\rm{DVPP}}}}}}{{{\omega _n}}}
		\end{aligned} \right.
        \label{power-2}
	\end{equation}
	
	On this basis, we derive the active power injections under multi-disturbance in (\ref{power-multi})-(\ref{power-5}). From (\ref{power-1})-(\ref{power-5}), it can be seen that the active power injections provided by the DVPP are actually determined by the aggregated virtual inertia~($H_\text{DVPP}$) and virtual damping~($D_\text{DVPP}$).
	\begin{equation}
		\Delta P^\text{Inj}(t) = \left\{ \begin{aligned}
			& \Delta P^\text{Inj}_{1}(t),\Delta t_{1} \le t < \tau + \Delta t_{2}\\ 
			& \Delta P^\text{Inj}_{2}(t),\tau + \Delta t_{2} \le t < 2\tau+\Delta t_{3}\\
			& \vdots \\
			& \Delta P^\text{Inj}_{n}(t),(n-1)\tau+\Delta t_{n} \le t \le n\tau\\
		\end{aligned} \right.
		\label{power-multi}
	\end{equation}
	where
	\begin{equation}
		\left\{ \begin{aligned}
			& \Delta P^\text{Inj}_{1}(t) = \mathcal{P}_{1}\cdot I\left[\Delta P_{1},t-\Delta t_{1} \right] \\
			& \Delta P^\text{Inj}_{2}(t) = \mathcal{P}_{2}\cdot I\left[\Delta P_{2}, t - (\tau+\Delta t_{2}) \right] + \Delta P^\text{Inj}_{1}(\tau+\Delta t_{2})\\
			& \vdots\\
			& \Delta P^\text{Inj}_{n}(t) = \mathcal{P}_{n}\cdot I\left[\Delta P_{n}, t-((n-1)\tau+\Delta t_{n}) \right] \\
            & + \Delta {P_{n - 1}^\text{Inj}}\left[ (n-1)\tau+\Delta t_{n} \right]
		\end{aligned} \right.
		\label{power-5}
	\end{equation}

    There are two types of frequency events: a frequency drop caused by a negative active power disturbance ($\Delta P < 0$) and a frequency increase caused by a positive disturbance ($\Delta P > 0$). The former results from an increase in load or a decrease in generation, while the latter is the opposite. In the case of $\Delta P < 0$, the DVPP participates in frequency regulation by providing active power injections from IBRs to support the frequency and ensure it remains within a secure range. In contrast, for $\Delta P > 0$, the DVPP compensates for the active power disturbance by reducing the output power of the IBRs to the grid.
    
\subsection{Two-Stage Required Parameters Decision-Making}

    In this section, we first analyze the steady-state performance of $\Delta P^\text{Inj}(t)$. To maintain the steady-state value of the frequency regulation power $\Delta P^\text{Inj}(t)$ at a low level, we minimize the steady-state change in frequency regulation power after each disturbance, as derived in (\ref{power-ss}). Therefore, the value of aggregated virtual damping~($D_\text{DVPP}$) should be adjusted to its minimum, satisfying the problem formulated in (\ref{power-D}).
    \begin{equation}
        \mathcal{S}(\Delta P_{i},D_\text{DVPP}) = \frac{\Delta P_{i}}{1+\frac{D_{0}+R}{D_\text{DVPP}}}    
        \label{power-ss}
    \end{equation}	
    \begin{equation}
		\begin{aligned}
			& D_{{\rm{DVPP}}}^\text{re} = \arg \min \mathcal{S}(\Delta P_{i},D_\text{DVPP})\\
			& \text{s.t.}\\
			& D_\text{DVPP}^\text{re} \in \Omega(H_\text{DVPP},D_\text{DVPP})
		\end{aligned}
		\label{power-D}
	\end{equation}

    After determining the required virtual damping~($D_\text{DVPP}^\text{re}$) in (\ref{power-D}), the aggregated virtual inertia~($H_\text{DVPP}$) is determined by considering the dynamic performance of the active power response, specifically the decay rate described in~(\ref{power-dynamic}) according to (\ref{power-1})-(\ref{power-2}). Thus, the value of virtual inertia should be adjusted to its minimum within the feasible region defined in (\ref{safety-region}) to achieve a lower overshoot or higher decay rate of the active power, as outlined in (\ref{power-H}).
    \begin{equation}                                                                             \zeta {\omega _n} = \frac{{2H + D{T^{{\rm{SG}}}}}}{{4H{T^{{\rm{SG}}}}}} = \frac{1}{{2{T^{{\rm{SG}}}}}} + \frac{D}{{4H}}
	\label{power-dynamic}
	\end{equation}
    
    \begin{equation}
		\begin{aligned}
			& H_{{\rm{DVPP}}}^\text{re} = \arg \min \zeta \omega_{n}(H_\text{DVPP})\\
			& \text{s.t.}\\
			& H_\text{DVPP}^\text{re} \in \Omega(H_\text{DVPP},D_\text{DVPP}^\text{re})
		\end{aligned}
		\label{power-H}
	\end{equation}
    
	\subsection{Regulation Power Reserve Decision-Making}

    By determining the frequency requirements, i.e., $H^\text{re}_\text{DVPP}$ and $D^\text{re}_\text{DVPP}$, at the DVPP level, the specific parameters of IBRs, i.e., $H_{i}$ and $D_{i}$, can be determined through optimal allocation. With various economic preferences of IBRs, we establish the parameter optimization model within DVPP in (\ref{allocation}).
	\begin{equation}
		\begin{aligned}
			& \mathop {\min }\limits_{{H_i},{D_i}} \sum\nolimits_{i = 1}^N {\left( {{a_i}{H_i} + {b_i}{D_i}} \right)}\\
			& \text{s.t.}\\
			& \sum\nolimits_{i}^{N}{{{H}_{i}}}={{H}_\text{DVPP}^\text{re}}, \\
			& \sum\nolimits_{i}^{N}{{{D}_{i}}}={{D}_\text{DVPP}^\text{re}}, \\
            & \left| \Delta P^\text{IBR}_{i} \right| \le \Delta P^\text{av}_{i},\ \ i = 1,2,...,N,\\
			& {H_{\min }} \le {H_i} \le {H_{\max }},\ i = 1,2,...,N,\\
			& {D_{\min }} \le {D_i} \le {D_{\max }},\ i = 1,2,...,N
		\end{aligned}
		\label{allocation}
	\end{equation}
	where $H_{i}$ and $D_{i}$ are the controller parameters, i.e., virtual inertia and virtual damping, of the $i$th IBR and $N$ is the number of IBRs within the DVPP. Besides, $a_{i}$ and $b_{i}$ are the costs of inertia and damping provision of the $i$th IBR and the corresponding active power injection is $\Delta P^\text{av}_{i}$. The $\Delta P_{i}^\text{IBR}$ is a piecewise function which can be proved as a linear combination of decision parameters $H_i$ and $D_i$. The detailed derivation and mathematical proof are in the Appendix. Thus, the optimization problem formulated in~(\ref{allocation}) is convex and can guarantee the global optimum.

    After solving the optimization model in (\ref{allocation}), the specific control parameters for each IBR are determined. On this basis, the power reserve during the frequency regulation period can also be determined by calculating the corresponding upward and downward peak values of regulation power responses of IBRs ($\Delta P^\text{IBR}_{i}(t),i=1,2,..., N$), which is formulated completely in Appendix. Thus, the optimal power reserve for frequency regulation is derived in (\ref{rev_upper})-(\ref{rev_down}), containing the upward power reserve~($R^{U}$) and downward power reserve~($R^{D}$), with the frequency security metrics being satisfied. This regulation power reserve decision-making process highlights the capability of the DVPP in energy management, which thoroughly considers the economic diversity of IBRs~\cite{ADD-2,ADD-3}. 
    
    To summarize, this paper focuses on constructing a robust feasible region for the calculation and allocation of frequency regulation power reserves, while ensuring the closed-loop stability of the system. The flowchart of the proposed robust reserve decision-making approach, along with the corresponding sections of the manuscript, is presented in Fig.~\ref{fig.flowchart}. As depicted in Fig.~\ref{fig.flowchart}, the process is divided into three stages. The first two stages involve identifying the worst disturbance and formulating the robust parameter feasible region, with minimal computational burden, as these are algebraic operations based on closed-form formulations. Uncertainty is handled through forecasting and real-time measurement. The third stage involves selecting virtual inertia and damping parameters through optimization, which can be accelerated by discretizing the feasible region. As the optimization is a convex program, it ensures a low computational load. Therefore, the approach has the potential to be effectively implemented in real-time.

    It should be pointed out that the proposed approach mainly focuses on the uncertainty from sequential disturbances, which pose a dominant challenge to frequency security in low-inertia power systems. Furthermore, the proposed approach can adaptively adjust its scheduling strategy based on real-time measurements of the system parameters, e.g, grid inertia ($H_0$) and damping ($D_0$). Moreover, unmodeled high-order dynamics beyond the active power–frequency deviation relationship are neglected for analytical tractability, which will be addressed in future work.

    \begin{figure*}
        \begin{equation}
		R^{U} = \left\{
        \begin{aligned}
            &\sum_{i=1}^{N} \left( \max_{0\le t \le n\tau}\Delta P^\text{IBR}_{i}(t) \right),\ \sum_{i=1}^{N} \left( \max_{0\le t \le n\tau}\Delta P^\text{IBR}_{i}(t) \right)\ge 0\\
            &0,\ \sum_{i=1}^{N} \left( \max_{0\le t \le n\tau}\Delta P^\text{IBR}_{i}(t) \right) <0
        \end{aligned}
        \right.
		\label{rev_upper}
	\end{equation}
    \begin{equation}
		R^{D} = \left\{
        \begin{aligned}
            &\sum_{i=1}^{N} \left( \min_{0\le t \le n\tau}\Delta P^\text{IBR}_{i}(t) \right),\ \sum_{i=1}^{N} \left( \min_{0\le t \le n\tau}\Delta P^\text{IBR}_{i}(t) \right)\le 0\\
            &0,\ \sum_{i=1}^{N} \left( \min_{0\le t \le n\tau}\Delta P^\text{IBR}_{i}(t) \right) > 0
        \end{aligned}
        \right.
		\label{rev_down}
	\end{equation}
    \hrulefill
    \end{figure*}

    \begin{figure}
		\centering
		\includegraphics[width=1\linewidth]{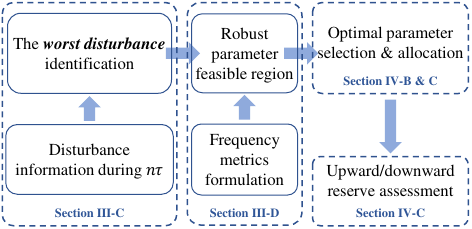}
		\caption{The flowchart of the proposed robust reserve decision-making approach.} 
		\label{fig.flowchart}
	\end{figure}
    
	\section{Case Studies} \label{sec5}
	
	\subsection{Set Up}

	\begin{figure}
		\centering
		\includegraphics[width=1\linewidth]{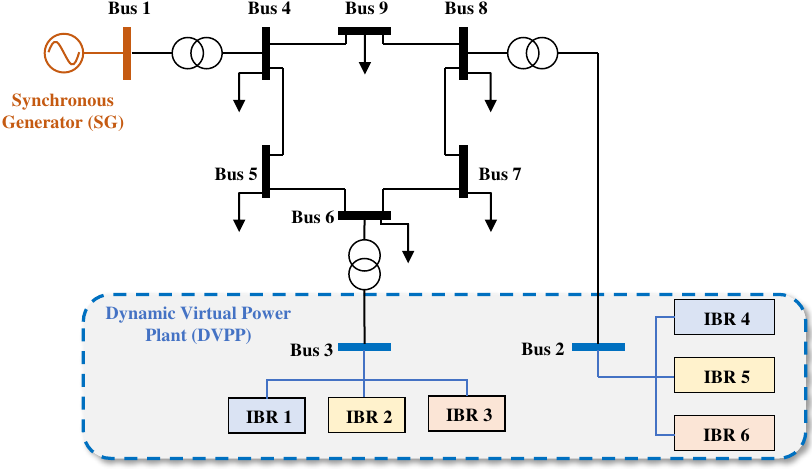}
		\caption{The diagram of the modified nine-bus systems used for case studies~\cite{Case-1}.} 
		\label{fig.39bus}
	\end{figure}
	
	The proposed approach is validated using a modified IEEE nine-bus system, which includes one synchronous generator (SG) with a capacity of 135 MW and one DVPP comprising six IBRs (a total of 140 MW) capable of providing the required active power injections for frequency regulation (Fig. \ref{fig.39bus})~\cite{Case-1}. The SG and DVPP are connected to the system via bus 1 and buses 2 and 3, respectively. The active power disturbances are introduced at bus 5 through load addition and removal. 

    According to the assumptions in Section~\ref {sec3}, the case study assumes that the disturbance magnitudes can be obtained through accurately forecasting the system load and power generation, which are set as 0.095 (p.u.), 0.109 (p.u.), -0.204 (p.u.), -0.158 (p.u.), and 0.253 (p.u.), respectively. The probabilities of individual disturbances are set to 0.8, 0.5, 0.8, 0.9, and 0.7, respectively. These values are empirically chosen to reflect the high frequency of short-term disturbances reported in recent literature~\cite{LoadFlu-2,Flu1,Flu2}. Although selected empirically, the proposed method ensures robustness regardless of the specific $\mathcal{P}_{i}$ values. 
    
    Additionally, for analysis, a disturbance occurrence point is set every 0.5$\tau$ (with uniform distribution of probabilities), resulting in a total of 10 points (with a regulation duration of 10$\tau$). The cost coefficients for inertia ($a_i$) and damping ($b_i$) are set to ensure that the dynamic support costs are reasonable and comparable to those of conventional ancillary services, as analyzed in earlier literature~\cite{COM1}. The other detailed parameter values used in the case studies are provided in TABLE~\ref{tab:parameter}.
    
    Each IBR adjusts its active power injections according to the determined control parameters. The above control parameters through the power outer loop influence reference set points of current~\cite{Outloop-1}. The simulations are conducted using MATLAB/Simulink R2024b, while the allocation optimization is implemented using Gurobi 10.0.1 on a desktop equipped with an Intel Core i7-10700 2.90 GHz CPU and 32 GB RAM. 
	
	\begin{table}[htbp]
		\centering
		\caption{Parameters for case studies}
		\begin{tabular}{ccc}
			\toprule
			Parameter & Value & Unit \\
			\midrule
			$D_{0}$     & 2     & MW/Hz \\
			$H_{0}$     & 10    & MWs/Hz \\
			$R$     & 10    & MW/Hz \\
			$T^\text{SG}$     & 7     & s \\
            $n$     & 5      & / \\
            $\tau$  & 60     & s \\
			$a_{i}$     & 3,4,1,1,2,1 & \$/(s) \\
            $b_{i}$     & 2,3,1,1,1.6,1 & \$/(p.u.) \\
			$P_{i}^\text{av}$     & 12.12,10.9,1.22,2.42,8.48,4.84 & MW \\
			$\Delta f_\text{lim}^\text{RoCoF}$    & 0.4   & Hz/s \\
			$\Delta f_\text{lim}^\text{Nadir}$    & 0.55   & Hz \\
            $\Delta f_\text{lim}^\text{ss}$    & 0.45   & Hz \\
			${H}_{\min},{H}_{\max}$    & 0.1, 6 & MWs/Hz \\
			${D}_{\min},{D}_{\max}$    & 0.1, 6 & MW/Hz \\
			\bottomrule
		\end{tabular}
		\label{tab:parameter}
	\end{table}

    \subsection{The Worst Disturbance Assessment}
    
    Based on previous analysis, the worst disturbances refer to the cases where all possible disturbances occur and include the maximum possible magnitude. For convenience, the sequential disturbance is described as a vector $\left[ \Delta t_{1}, \Delta t_{2}, \Delta t_{3}, \Delta t_{4}, \Delta t_{5} \right]$ where the elements represent the occurrence of each disturbance~(as shown in Fig~\ref{fig.timeaxis}).

    Based on the setup in the previous section, the worst disturbance is identified as $ \left[ \tau, \tau, 3\tau, 3\tau, \chi(4\tau,5\tau) \right]$ where $\chi(4\tau,5\tau)$ means any time points between $4\tau$ and $5\tau$. Because when disturbances with the same sign occur at the same time, the combined disturbance will result in a higher RoCoF and frequency nadir. It should be noted that the worst disturbances are also not unique due to the non-uniqueness of the value of $\chi(4\tau,5\tau)$. However, their impacts on RoCoF and frequency nadir are the same. One of these cases $L^{w} =\left[ \tau, \tau, 3\tau, 3\tau, 4\tau \right]$ will be selected for the calculation in the following.
    
    \subsection{Parameter Determined and Verification} 
    
    As for the case with the worst disturbances ($L^{w}$) being analyzed before, its robust feasible parameter region~($\Omega$) can be calculated according to~(\ref{safety-region}). According to the analysis in Section III. D, the feasible region satisfies the constraints on RoCoF~($\Delta f_\text{lim}^\text{RoCoF}$, calculation based on existing standards~\cite{Std-IEEE}), nadir~($\Delta f_\text{lim}^\text{Nadir}$) and steady-state values~($\Delta f_\text{lim}^\text{ss}$).
    
    \begin{table}[htbp]
        \centering
        \caption{Three parameter combinations for comparison}
        \begin{tabular}{ccc}
            \toprule
            Parameter Combination & $H_\text{DVPP}$ (s) & $D_\text{DVPP}$ (p.u.) \\
            \midrule
             1~(for comparison)   & 10   & 8 \\
             2~(for comparison)   & 14 & 10.68 \\
             3~(optimal)          & 19.86  & 10.68 \\
            \bottomrule
        \end{tabular}%
        \label{tab1}%
    \end{table}%  

    \begin{figure}
	   \centering
	   \includegraphics[width=0.95\linewidth]{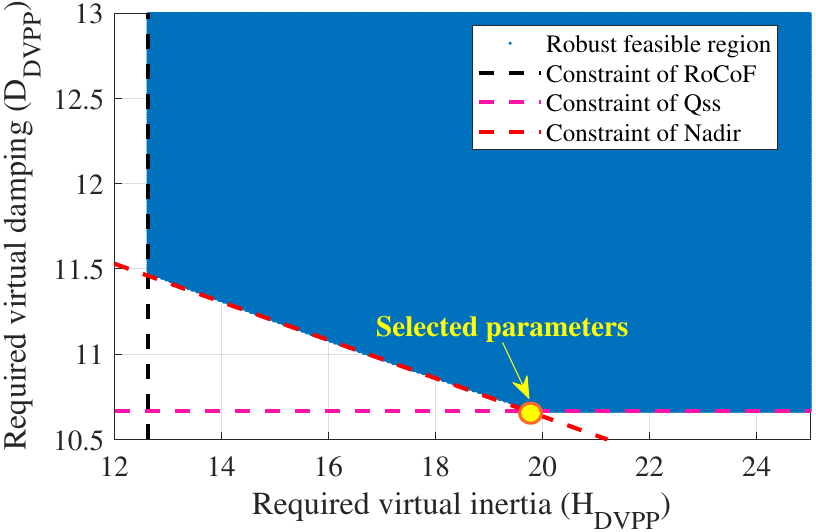}
	   \caption{The robust feasible parameter region of the aggregated virtual inertia and virtual damping for the DVPP.} 
	   \label{fig.region-1}
    \end{figure}
    
\begin{figure}
	   \centering
	   \includegraphics[width=0.95\linewidth]{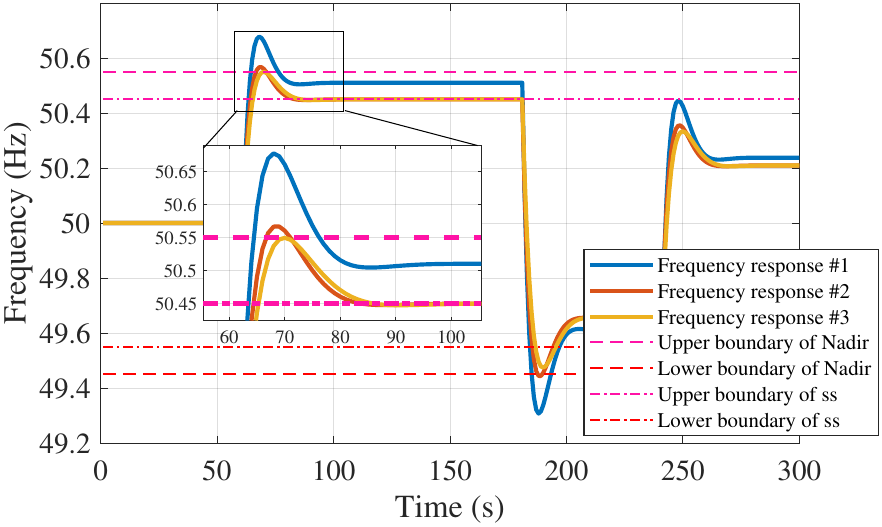}
	   \caption{The frequency responses of center of inertia~(COI) under three parameter combinations.} 
	   \label{fig.Com-1}
    \end{figure}

\begin{figure}
	   \centering
	   \includegraphics[width=0.95\linewidth]{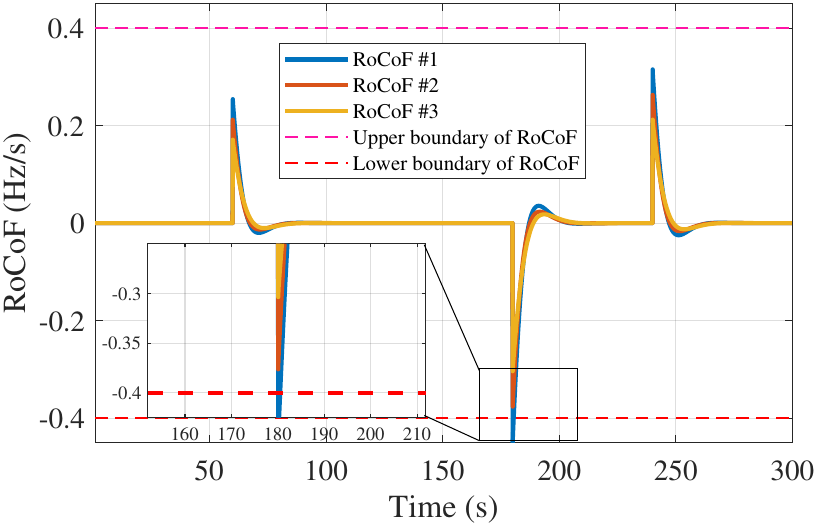}
	   \caption{The RoCoF of COI under three parameter combinations.} 
	   \label{fig.Com-2}
    \end{figure}

\begin{figure}
	   \centering
	   \subfloat[Frequency response]{
 		\includegraphics[width=0.48\linewidth]{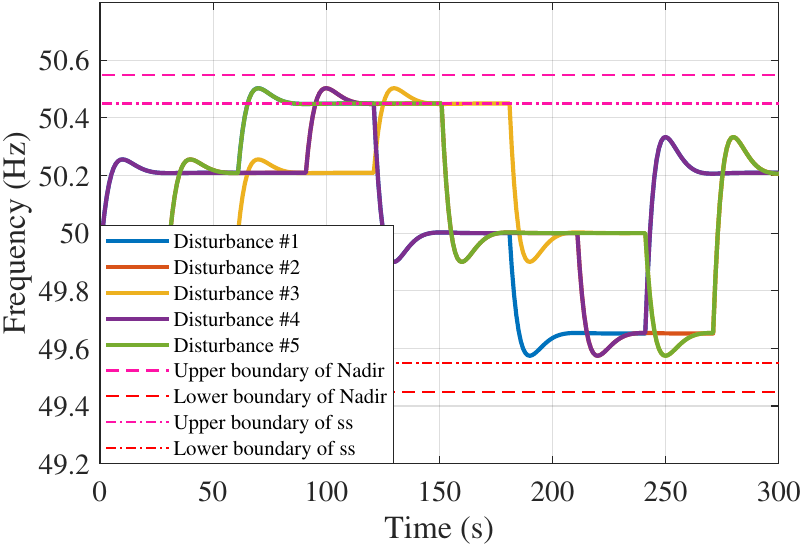}}
          \subfloat[RoCoF]{
 		\includegraphics[width=0.48\linewidth]{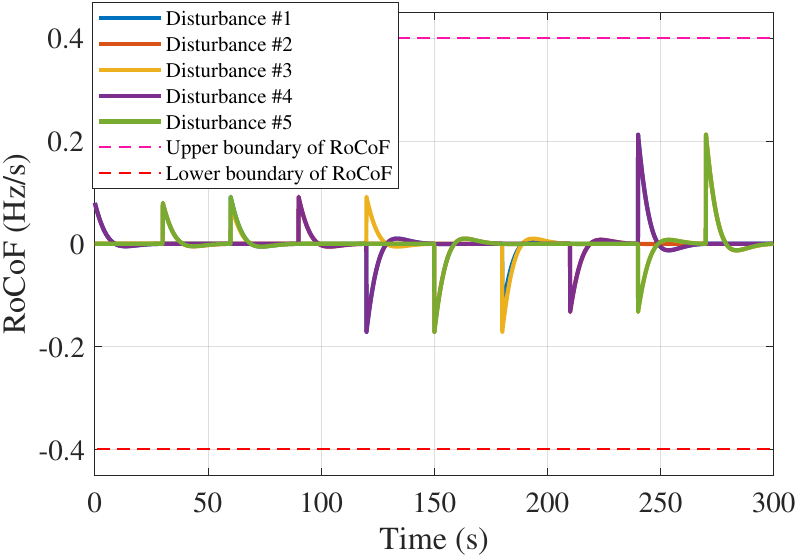}}
	   \caption{The frequency responses of COI under five disturbances with $H_\text{DVPP}^\text{re}=19.86$, $D_\text{DVPP}^\text{re}=10.68$.} 
	   \label{fig.Com-3}
    \end{figure}

     The parameter region visualization is shown in Fig.~\ref{fig.region-1} where the parameter combination ($H_\text{DVPP}^\text{re}=19.86$ s, $D_\text{DVPP}^\text{re}=10.68$ p.u.) is determined based on the analysis in Section~\ref{sec4}. B. With the selected parameters, the DVPP can provide sufficient support for the system frequency. Fig.~\ref{fig.Com-1}-Fig.~\ref{fig.Com-2} present the comparison among the selected parameters and two-parameter combinations listed in Tab.~\ref{tab1}. From the comparison results of the nadir criterion, the parameter combinations selected based on the proposed method~(the no. 3 combination) yield frequency dynamics that satisfy the preset upper and lower boundaries of three security metrics. In contrast, the other two parameter combinations outside the feasible parameter domain result in frequency dynamics that cannot simultaneously satisfy all three security constraints. In addition to $L^{w}$, we selected five other typical disturbance scenarios~(listed in Tab.~\ref{tab2}) and tested the effectiveness of the chosen parameters under each scenario. As shown in Fig.~\ref{fig.Com-3}, the selected parameters ensure frequency security under different disturbance scenarios.

    \begin{table}[htbp]
        \centering
        \caption{Five disturbance scenarios with varying disturbance times for comparison}
        \begin{tabular}{cccccc}
            \toprule
            Disturbance & \multicolumn{1}{l}{$\Delta P_{1}$} & \multicolumn{1}{l}{$\Delta P_{2}$} & \multicolumn{1}{l}{$\Delta P_{3}$} & \multicolumn{1}{l}{$\Delta P_{4}$} & \multicolumn{1}{l}{$\Delta P_{5}$} \\
            \midrule
            1     & 0     & $\tau$     & 2$\tau$     & 3$\tau$     & 4$\tau$ \\
            2     & 0.5$\tau$   & 1.5$\tau$   & 2.5$\tau$   & 3.5$\tau$   & 4.5$\tau$ \\
            3     & 1$\tau$     & 2$\tau$     & 3$\tau$     & 4$\tau$     & 4.5$\tau$ \\
            4     & 0     & 1.5$\tau$   & 2$\tau$     & 3.5$\tau$   & 4$\tau$ \\
            5     & 0.5$\tau$   & 1 $\tau$    & 2.5$\tau$   & 4$\tau$     & 4.5$\tau$ \\
            \bottomrule
        \end{tabular}%
        \label{tab2}%
    \end{table}%
    
\begin{figure}
	   \centering
	   \subfloat[Frequency response]{
 		\includegraphics[width=0.48\linewidth]{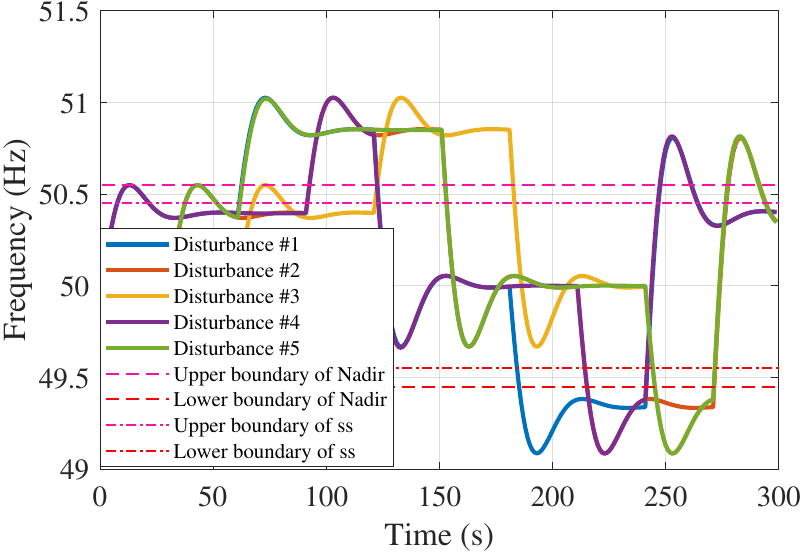}}
          \subfloat[RoCoF]{
 		\includegraphics[width=0.48\linewidth]{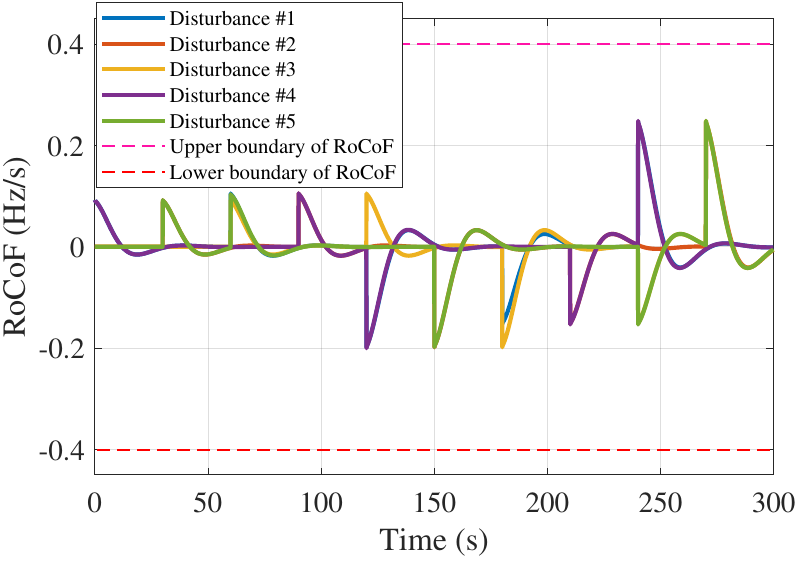}}
	   \caption{The frequency responses of COI under five disturbances with $H_\text{DVPP}^\text{re}=16.05$, $D_\text{DVPP}^\text{re}=0$ obtained from the superposition of disturbances~\cite{COM1, COM2}.} 
	   \label{fig.Com-4}
    \end{figure}

    The traditional research approach~\cite{COM1, COM2} involves superimposing multiple disturbances that may occur within the regulation period onto the same point (usually taken as the starting point) for simplified analysis. Using this method, the results yield $H_\text{DVPP}^\text{re}=16.05$ s, $D_\text{DVPP}^\text{re}=0$ p.u., which is designed as the baseline for comparison. Under the same disturbances listed in Tab.~\ref{tab2}, the corresponding frequency dynamics pose risks of exceeding the nadir and steady-state limits under possible disturbance scenarios, as shown in Fig.~\ref{fig.Com-4}. Because the algebraic superposition cancels out the correlation between adjacent disturbances, leading to overly non-conservative results. 

	\subsection{Power Reserve Calculation and Allocation for DVPP}

\begin{figure}
	   \centering
	   \includegraphics[width=0.95\linewidth]{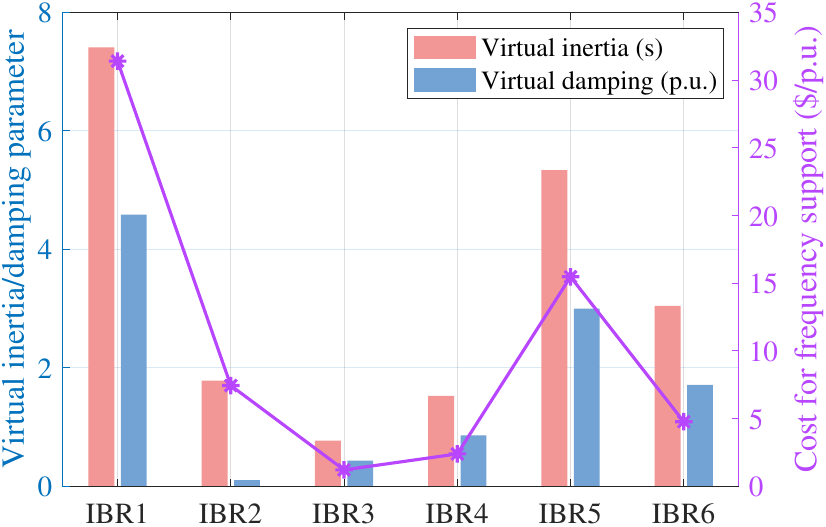}
	   \caption{The allocation results of virtual inertia/damping parameters for IBRs.} 
	   \label{fig.Allocation}
    \end{figure}

\begin{figure}
	   \centering
	   \includegraphics[width=0.95\linewidth]{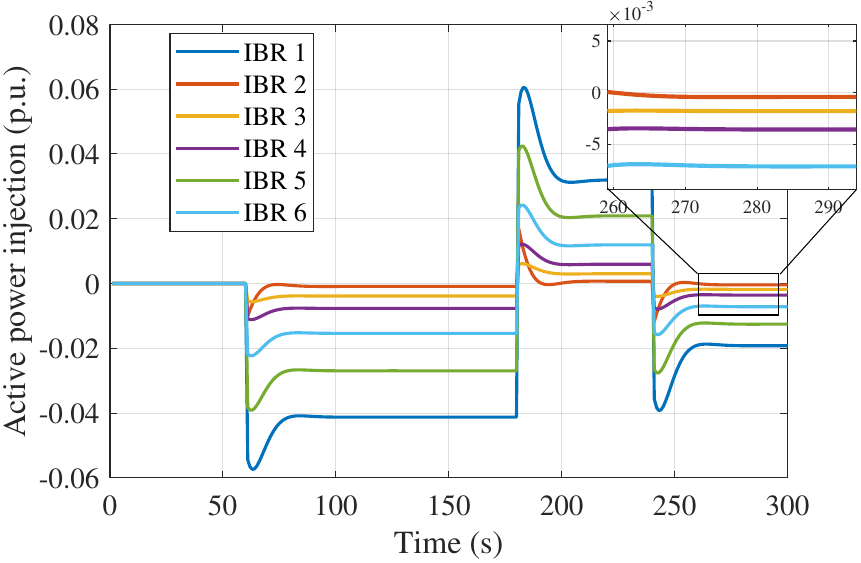}
	   \caption{The active power injection of IBRs during the frequency regulation period.} 
	   \label{fig.power}
    \end{figure}

    The optimal allocation model (\ref{allocation}) is a convex optimization problem that can be solved directly using commercial solvers, such as Gurobi, as utilized in this paper. The results are shown in Fig.~\ref{fig.Allocation}. The allocation of virtual inertia parameters ($H_{i}$) and virtual damping parameters ($D_{i}$) varies across the six IBRs due to their economic and capacity differences. After optimization, the total frequency support cost, achieved by adjusting the parameters for each IBR, is \$62.624. Additionally, under $L^{w}$, the active power injections for the IBRs are presented in Fig.~\ref{fig.power}, with the total upward and downward reserves of the DVPP being 0.1623 (p.u.) and 0.1473 (p.u.), respectively. 

    Based on the allocation results presented in Fig.~\ref{fig.Allocation}-\ref{fig.power}, the variation in the active power injection of the IBRs depends on the adjustable capacity of the IBRs and their respective differentiated adjustment costs. This is the result of a cost-optimization trade-off. Overall, IBR 1, 5, and 6 are allocated more regulating power due to their larger adjustable capacities, which together account for 64\% of the total capacity. In contrast, the frequency regulation power of IBR 2 is relatively low because its adjustment cost is too high, despite its adjustable capacity accounting for 27\% of the total.

    \subsection{Discussion}

    The above case study analysis is based on disturbance analysis with adjacent disturbances having the same direction, which is one type. Another possible type involves adjacent disturbances in opposite directions. The key difference is that, when analyzing the worst disturbance for the latter case, adjacent disturbances must be prevented from superimposing. However, the analysis and calculation methods for this case are similar to those described earlier.

    Additionally, this paper addresses frequency support optimization under disturbances with uncertain occurrence times by seeking the worst disturbance. While this approach is conservative, the degree of conservatism depends on the probability of the worst disturbance occurring at any given time. However, in comparison to the power system's frequency safety, the economic losses due to such conservative decision-making are not the primary concern in this paper.
    
	\section{Conclusion} \label{sec6}

    This paper presents a robust reserve decision-making approach for the DVPP, which aggregates numerous IBRs to provide frequency support under temporally sequential and uncertain disturbances. By fully capturing the temporal dependencies of these disturbances, the relationship between system frequency response and regulation power of the DVPP is analytically derived. Based on this, we ensure system frequency security through a sequential process involving worst disturbance identification, formulation of a robust parameter feasible region, and selection of tunable parameters. Once the parameters, including virtual inertia and virtual damping for the IBRs, are determined, the corresponding regulation power reserves are calculated. Case studies suggest that the robust parameter feasible region is sufficient for guaranteeing frequency security on RoCoF, nadir and steady-state deviation metrics. Additionally, compared to the traditional single-step simplification scheme, the parameter decisions proposed in this paper reduce RoCoF, frequency nadir, and steady-state deviation by 66.08\%, 44.36\%, and 32.74\%, respectively, demonstrating greater reliability in ensuring frequency security.
    
    Future work will focus on more detailed system modeling, specifically addressing unmodeled high-order dynamics and the dynamic behavior under overlapping disturbance events and frequency response processes. Additionally, the secondary frequency control will be explicitly incorporated into a multi-timescale framework to examine its interactions with the faster inertia response and primary frequency control.
    
	\appendix
	\setcounter{equation}{0}
	\renewcommand{\theequation}{A\arabic{equation}}
    The regulation power dynamics of $i$th IBR is formulated as in (\ref{app-1})-(\ref{app-4}), where $H_{i}$ and $D_{i}$ are the power outer loop parameters of the $i$th IBR.
    \begin{equation}
		\Delta P^\text{IBR}_{i}(t) = \left\{ \begin{aligned}
			& \Delta P^\text{IBR}_{i,1}(t),\Delta t_{1} \le t < \tau + \Delta t_{2}\\ 
			& \Delta P^\text{IBR}_{i,2}(t),\tau + \Delta t_{2} \le t < 2\tau+\Delta t_{3}\\
			& \vdots \\
			& \Delta P^\text{IBR}_{i,n}(t),(n-1)\tau+\Delta t_{n} \le t \le n\tau\\
		\end{aligned} \right.
		\label{app-1}
	\end{equation}
	where
	\begin{equation}
		\left\{ \begin{aligned}
			& \Delta P^\text{IBR}_{i,1}(t) = \mathcal{P}_{1}\cdot \mathbb I_{i}\left[\Delta P_{1},t-\Delta t_{1} \right] \\
			& \Delta P^\text{IBR}_{i,2}(t) = \mathcal{P}_{2}\cdot \mathbb I_{i}\left[\Delta P_{2}, t - (\tau+\Delta t_{2}) \right] + \Delta P_{1}^\text{IBR}(\tau+\Delta t_{2})\\
			& \vdots\\
			& \Delta P^\text{IBR}_{i,n}(t) = \mathcal{P}_{n}\cdot \mathbb I_{i}\left[\Delta P_{n}, t-((n-1)\tau+\Delta t_{n}) \right] \\
            & + \Delta {P_{n - 1}^\text{IBR}}\left[ (n-1)\tau+\Delta t_{n} \right]
		\end{aligned} \right.
		\label{app-2}
	\end{equation}

   \begin{equation}
		\mathbb I_{i}(\Delta P,t) =  - \frac{{\mathbb A_{i} + {e^{ - \zeta {\omega _n}t}}\left( {\frac{{\mathbb C_{i} - \zeta {\omega _n}\mathbb B_{i}}}{{{\omega _d}}}\sin ({\omega _d}t) + \mathbb B_{i}\cos ({\omega _d}t)} \right)}}{{2H{T^{{\rm{SG}}}}}}
		\label{app-3}
	\end{equation}
	\begin{equation}
		\left\{ \begin{aligned}
			& \mathbb A_{i} = \frac{{\Delta P \cdot {D_{{i}}}}}{{\omega _n^2}}\\
			& \mathbb B_{i} = 2{T^{{\rm{SG}}}}{H_{{i}}}\Delta P - \frac{{\Delta P \cdot {D_{{i}}}}}{{\omega _n^2}}\\
			& \mathbb C_{i} = ({T^{{\rm{SG}}}}{D_{{i}}} + 2{H_{{i}}})\Delta P - \frac{{2\zeta \Delta P \cdot {D_{{i}}}}}{{{\omega _n}}}
		\end{aligned} \right.
        \label{app-4}
	\end{equation}

    The Eq.~(\ref{app-3}) can be rearranged as in (\ref{app-5}) which can be seen as a linear combination of inertia ($H_i$) and damping ($D_i$), where the intermediate variables $\alpha$ and $\beta$ are listed in (\ref{app-6})-(\ref{app-7}) and the variables depend on system-level parameters, i.e., $H_0$, $D_0$, $H_\text{DVPP}$, $H_\text{DVPP}$, $T^\text{SG}$, etc. In this way, variables $\alpha$ and $\beta$ can be considered as constants for $H_i$ and $D_i$.

    \begin{equation}
    \begin{aligned}
        &{{\mathbb{I}}_i}(\Delta P,t) =  - \frac{{{{\mathbb{A}}_i} + {e^{ - \zeta {\omega _n}t}}\left( {\frac{{{{\rm{C}}_i} - \zeta {\omega _n}{{\mathbb{B}}_i}}}{{{\omega _d}}}\sin \left( {{\omega _d}t} \right) + {{\mathbb{B}}_i}\cos \left( {{\omega _d}t} \right)} \right)}}{{2H{T^{{\rm{SG}}}}}}\\
        & = \alpha {H_i} + \beta {D_i}
    \end{aligned}
    \label{app-5}
    \end{equation}

    \begin{equation}
        \begin{aligned}
            &\alpha  = \left( {\zeta {\omega _n} - \frac{{{e^{ - \zeta {\omega _n}t}}\cos \left( {{\omega _d}t} \right)}}{{2H{T^{{\rm{SG}}}}}}} \right)2{T^{SG}}\Delta P \\
            &- 2\frac{{{e^{ - \zeta {\omega _n}t}}\sin \left( {{\omega _d}t} \right)}}{{2H{T^{{\rm{SG}}}}{\omega _d}}}\Delta P
        \end{aligned}
        \label{app-6}
    \end{equation}

    \begin{equation}
        \begin{aligned}
            &\beta  =  - \frac{{\Delta P}}{{2H{T^{{\rm{SG}}}}\omega _n^2}} - \frac{{{e^{ - \zeta {\omega _n}t}}\sin \left( {{\omega _d}t} \right)}}{{2H{T^{{\rm{SG}}}}{\omega _d}}}\left( {{T^{{\rm{SG}}}}\Delta P - \frac{{2\zeta \Delta P}}{{{\omega _n}}}} \right)\\
            &- \left( {\zeta {\omega _n} - \frac{{{e^{ - \zeta {\omega _n}t}}\cos \left( {{\omega _d}t} \right)}}{{2H{T^{{\rm{SG}}}}}}} \right)\frac{{\Delta P}}{{\omega _n^2}}
        \end{aligned}
        \label{app-7}
    \end{equation}

    Because $\Delta P_i^{{\rm{IBR}}}(t)$ is a linear combination of ${{\mathbb{I}}_i}(\Delta P,t)$, thus $\Delta P_i^{{\rm{IBR}}}(t)$ can also be formulated as a linear combination of $H_i$ and $D_i$, which are the decision variables of program (\ref{allocation}). In summary, the optimization problem (\ref{allocation}) is a convex program of $H_i$ and $D_i$.
    
	\footnotesize
	\bibliography{reference}
	\bibliographystyle{IEEEtran}

\end{document}